\documentclass[aoas,MSNbibl,nameyear,dvips]{arximspdf}
\usepackage{graphicx,url,breakurl}
\doi{10.1214/14-AOAS723} %kopijuoti is PTS
\volume{8}
\issue{2}
\pubyear{2014}
\firstpage{905}
\lastpage{925}


\begin{document}
\begin{frontmatter}

\title{Pairwise comparison of treatment levels in functional analysis
of variance with application to erythrocyte hemolysis}
\runtitle{A follow-up test for functional linear models}

\begin{aug}
\author[a]{\fnms{Olga} \snm{Vsevolozhskaya}\corref{}\ead[label=e1]{vsevoloz@math.montana.edu}\thanksref{m1}},
\author[a]{\fnms{Mark} \snm{Greenwood}\ead[label=e2]{greenwood@math.montana.edu}\thanksref{m1}}
\and\break
\author[b]{\fnms{Dmitri} \snm{Holodov}\thanksref{m2}\ead[label=e3]{holodov@pharm.vsu.ru}}
\runauthor{O. Vsevolozhskaya, M. Greenwood and D. Holodov}
\affiliation{Montana State University\thanksmark{m1} and Voronezh State
University\thanksmark{m2}}
\address[a]{O. Vsevolozhskaya\\
M. Greenwood\\
Department of Mathematical Sciences\\
Montana State University\\
P.O. Box 172400\\
Bozeman, Montana 59717-2400\\
USA\\
\printead{e1}\\
\phantom{E-mail:\ }\printead*{e2}} %adresu isvedimo komanda gale!
%Montana State University\\
\address[b]{D. Holodov\\
Department of Pharmaceutical Sciences\\
Voronezh State University\\
3 Studenchiskaya St.\\
Voronezh 394620\\
Russia\\
\printead{e3}}
\end{aug}

% HISTORY:
\received{\smonth{4} \syear{2013}}
\revised{\smonth{10} \syear{2013}}

% ABSTRACT
%
\begin{abstract}
Motivated by a practical need for the comparison of hemolysis curves at
various treatment levels, we propose a novel method for pairwise
comparison of mean functional responses. The hemolysis curves---the
percent hemolysis as a function of time---of mice erythrocytes (red
blood cells) by hydrochloric acid have been measured among different
treatment levels. This data set fits well within the functional data
analysis paradigm, in which a time series is considered as a
realization of the underlying stochastic process or a smooth curve.
Previous research has only provided methods for identifying some
differences in mean curves at different times. We propose a two-level
follow-up testing framework to allow comparisons of pairs of treatments
within regions of time where some difference among curves is
identified. The closure multiplicity adjustment method is used to
control the family-wise error rate of the proposed procedure.
\end{abstract}

% KEYWORDS
\begin{keyword}
\kwd{Functional data analysis}
\kwd{FANOVA}
\kwd{multiple comparison}
\kwd{permutation method}
\kwd{pairwise comparison}
\end{keyword}

\end{frontmatter}

%s1 #&#
\section{Introduction}\label{sec1}

The use of nonsteroidal anti-inflammatory drugs\break  (NSAIDs) is widespread
in the treatment of various rheumatic conditions [\citet{nasonov2006}].
Gastrointestinal symptoms are the most common adverse events associated
with the NSAID therapy [\citet{garcia2001}]. \citet{holodov2012}
suggested oral administration of a procaine (novocaine) solution in low
concentration (0.25 to 1\%) to reduce the risk of upper
gastrointestinal ulcer bleeding associated with NSAIDs. To validate the
effectiveness of the proposed therapy, an experiment was conducted to
study the effect of novocaine on the resistance of the red blood cells
(erythrocytes) to hemolysis by hydrochloric acid as well as efficacy of
novocaine dosage. Hydrochloric acid is a major component of gastric
juice and a lower rate of erythrocyte hemolysis should indicate a
protective effect of novocaine.

\begin{figure}

\includegraphics{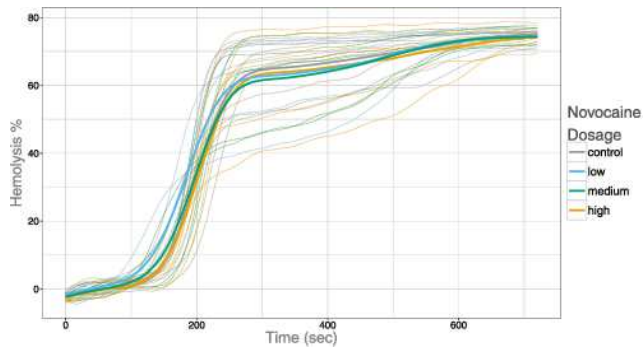}

\caption{Twenty hemolysis curves (erythrograms) of mice erythrocytes by
hydrochloric acid with superimposed estimated mean functions.}
\label{example}
\end{figure}

Hemolytic stability of erythrocytes for the control and for three
different dosages of novocaine ($4.9\times10^{-6}$ mol/L, $1.0\times
10^{-5}$mol/L, and $2.01\times10^{-5}$mol/L) was measured as a
percentage of hemolyzed cells. The data for the analysis were curves of
hemolysis (erythrograms) that were measured as functions of time.
Figure~\ref{example} illustrates a sample of percent hemolysis curves.
The goal of the statistical analysis was to summarize the associated
evidence across time of the novocaine effect including performing
pairwise comparisons of novocaine dosages.

%f1 #&#

Most current approaches essentially evaluate differences among groups
of curves point-wise. These approaches treat data that are inherently
functional (e.g., hemolysis is a smooth function of time) as a finite
vector of observations over time. A typical point-wise approach is to
perform a one-way analysis of variance (ANOVA) test at each time
point. However, when testing is performed at a large number of points
simultaneously, the type I error rate is going to be inflated. \citet
{cox2008} carefully investigated this issue and proposed a method that
utilizes a point-wise ANOVA approach, while properly controlling the
type I error rate.

Alternatively, function-valued methods can be employed. A key advantage
of the functional approach over its close counterpart---the
multivariate approach---is that the former retains information of the
ordering and spacing of observations over time. By assuming that there
is a true underlying functional response for each subject,
function-valued methods explicitly incorporate information over time.
Thus, a~method is ``functional'' if it approximates a finite vector of
observations by a function (a nonparametric function is a typical
choice) and then builds a test statistic based on these functional estimates.

The functional analysis of variance (FANOVA) can be employed to perform
testing among $k$ groups of curves. The overall functional testing
methods, such as the functional $\mathcal{F}$ of \citet{shen2004} or the
functional $V_n$ of \citet{cuevas2004}, can be utilized to test for
associated evidence across the entire functional domain (across all
time). \citet{vsevolozhskaya2013} developed a method for inferences in a
FANOVA situation on subregions of the initial functional domain.
However, none of these methods [including the point-wise method of \citet
{cox2008}] allows for pairwise comparisons of functional means. Thus,
the challenge for the current analysis was to determine differences
among novocaine dosages within specific intervals of time, where
significant differences among hemolysis curves are present [these time
intervals can be identified based on the methods in \citet{vsevolozhskaya2013}].

In this paper, we introduce a new function-valued two-step procedure:
first, to detect regions in time of significant differences among mean
curves, and, second, to perform a pairwise comparison of treatment
levels within those regions. The approach utilizes two ideas: (i)
combining methods to map a test statistic of the individual hypotheses,
$H_1, \ldots, H_m$, to the global one, $\bigcap_{i=1}^mH_i$, and (ii) the
closure principle of \citet{marcus1976} to control the family-wise error
rate (FWER), the probability of at least one false rejection. The rest
of the article is organized in the following manner. We give an
overview of the FANOVA problem and the existing methods for
investigating the functional domain for regions where significant
differences occur. We discuss the proposed procedure for investigating
regions of time for significant differences and detail a computational
shortcut that allows isolation of individual significance even for a
large number of tests. We extend the proposed procedure to perform
pairwise comparisons of the treatment levels within identified
functional regions of statistical significance. The protective effect
of novocaine is demonstrated based on the different patterns between
groups detected in certain regions of time.

%s2 #&#
\section{Methods}\label{sec2}

Functional analysis of variance involves testing for some difference
among $k$ functional means. In functional data analysis, $t$ is used to
denote a real-valued variable (usually of time) and $y(t)$ denotes a
continuous outcome, which is a function of $t$. Then, the FANOVA model
is written as
%
%e1 #&#
\begin{equation}
\label{eq:1} y_{ij}(t)=\mu_{i}(t) + \varepsilon_{ij}(t),
\end{equation}
where $\mu_i(t)$ is the mean function of group $i$ at time $t$,
$i=1,\ldots,k$, $j$ indexes a functional response within a group,
$j=1,\ldots, n_i$, and $\varepsilon_{ij}(t)$ is the residual function.
Each $\varepsilon_{ij}(t)$ is assumed to be a mean zero and independent
Gaussian stochastic process. The FANOVA hypotheses are written as
\begin{eqnarray*}
&&H_0\dvtx \quad\mu_1(t)=\mu_2(t)=\cdots=
\mu_k(t),
\\
&& H_a\dvtx\quad  \mu_i(t) \neq\mu_{i'}(t)\qquad
\mbox{for at least one $t$ and } i \neq i'.
\end{eqnarray*}
The alternative hypothesis considers any difference anywhere in $t$
among $k$ population means of $y_{ij}(t)$.

In recent years two different general approaches have emerged to
perform the FANOVA test. In \citet{shen2004}, as well as many other
papers [see \citet{cuevas2004}, \citet{ramsay2009} and \citet
{cuesta2010}], a global test statistic has been developed to perform
the FANOVA test. The statistic is ``global'' because it is used to
detect differences anywhere in the entire functional domain (anywhere
in $t$). An alternative approach [\citet{ramsay2005} and \citet{cox2008}]
is to use a point-wise (or individual) test statistic to perform
inference across $t$, that is, identify specific regions of $t$ with
significant difference among functional means.

%s2.1 #&#
\subsection{``Global'' approach}\label{sec2.1}
Suppose the domain $ [a, b ]$ of functional responses can be
split into $m$ prespecified mutually exclusive and exhaustive intervals
such that $ [ a, b ]=\bigcup_{i=1}^m [ a_i,b_i  ]$. For
instance, in the novocaine experiment the researchers were interested
in the effect of novocaine during specific time intervals associated
with hemolysis of different erythrocyte populations: hemolysis of the
least stable population ($ [a_2,b_2 ] = 61\mbox{--}165$ sec), general
population ($ [a_3,b_3 ] = 166\mbox{--}240$ sec), and most stable
($ [a_4,b_4 ] = {}$over 240 sec). For each interval $ [a_i,
b_i  ]$, $i=1, \ldots, m$, an individual functional statistic of
\citet{shen2004}, $\mathcal{F}_i, i=1, \ldots, m$, can be calculated as
%
%e2 #&#
\begin{equation}
\label{eq:Fsf} \mathcal{F}_i  =  \frac{ \int_{ [a_i, b_i  ]}\sum_{j=1}^k
n_j(\hat{\mu}_j(t)-\hat{\mu}(t))^2\,dt/(k-1)}{ \int_{ [a_i, b_i
]}\sum_{j=1}^k\sum_{s=1}^n(y_{js}(t)-\hat{\mu}_j(t))^2\,dt/(n-k)},
\end{equation}
where $n$ is the total number of functional responses and $k$ is the
number of groups. The numerator of the $\mathcal{F}$ statistic accounts
for ``external'' variability among functional responses and the
denominator for the ``internal'' variability. \citet{cuevas2004} argue
that the null hypothesis should be rejected based on the measure of the
differences among groups, that is, the ``external'' variability. Hence,
\citet{cuevas2004} proposed a statistic $V_n$ based on the numerator of
$\mathcal{F}$:
%
%e3 #&#
\begin{equation}
\label{eq:Vn} V_n=\sum_{i<j}^k
n_i\bigl\Vert \hat{\mu}_i(t)-\hat{\mu}_j(t)
\bigr\Vert ^2,
\end{equation}
where $\Vert  \cdot\Vert $ is the $L_2$ norm calculated over the $ [a_i,
b_i ]$ interval. \citet{gower1999} also argue that in a permutation
setting a test can be based just on the numerator of the test
statistic. That is, if only the numerator of the functional $\mathcal
{F}$ is used, the changes to the test statistic are monotonic across
all permutations and, thus, probabilities obtained are identical to the
ones obtained from the original $\mathcal{F}$. \citet{delicado2007}
points out that for a balanced design, the numerator of the functional
$\mathcal{F}$ and $V_n$ differ by only a multiplicative constant,
reinforcing how they provide the same results in a permutation setting.
\citet{vsevolozhskaya2013} fully extended this testing approach by
allowing identification of the time interval, $ [a_i, b_i ]$,
$i=1, \ldots, m$, within the time domain, $ [a, b ]$, while
having proper control of at least one false rejection.

%s2.2 #&#
\subsection{Point-wise approach}\label{sec2.2}

Suppose that a set of smooth functional responses is evaluated on a
dense grid of points, $t_1, \ldots, t_m$. For instance, the percentage
of hemolyzed cells can be evaluated every second. \citet{cox2008}
propose a test for differences in the mean curves from several
populations, that is, perform functional analysis of variance, based on
these discretized functional responses. First, at each of the $m$
evaluation points, the regular one-way analysis of variance test
statistic, $F_i, i=1, \ldots, m$, is computed. For each test the
$p$-value is calculated based on the parametric $F$-distribution and
then the Westfall--Young randomization method [\citet{westfall1993}] is
applied to correct the $p$-values for multiplicity. The implementation
of the method can be found in the \texttt{multtest} [\citet{multtest}] R
package [\citet{R}].

Certain criticisms may be raised for both the \textit{``global''} and
the \textit{point-wise} approaches. First, the \textit{point-wise}
approach can determine regions of the functional domain with a
difference in the means, but there is no clear way to extend this
approach to determine which pairs of populations are different. Second,
for the \citet{cox2008} procedure, the $p$-value for the global test cannot be obtained, which is an undesirable property since the method
might be incoherent between the global and point-wise inference. The
\textit{global} approach does not provide the time-specific detail that
the point-wise methods provide and the subregion inferences in \citet
{vsevolozhskaya2013} require specification of the subregions which may
be arbitrarily defined in some applications. We suggest a procedure
that overcomes the majority of these issues. By using a combining
function along with the closure principle of \citet{marcus1976}, we are
able to obtain the $p$-value for the overall test as well as adjust the
individual $p$-values for multiplicity. Additionally, the proposed
procedure allows us to perform a pairwise comparison of the group's
functional means and therefore determine which populations show
evidence of differences in each time region. However, the proposed
procedure still requires prespecification of these time regions, which
in some applications can be vague.

%s2.3 #&#
\subsection{Proposed methodology}\label{sec2.3}

Once again, suppose the domain $ [a, b ]$ is split into $m$
prespecified mutually exclusive and exhaustive intervals. We propose to
use the numerator of the functional $\mathcal{F}$ as the test statistic
$T_i$, $i=1,\ldots, m$, for each $ [a_i,b_i ]$, and then
utilize a combining function to obtain the test statistic for the
entire $ [a,b ]$. Typical combining functions have the same
general form: the global statistic is defined as a weighted sum,
$T=\sum w_iT_i$, of the individual statistics with some $w_i$ weights
[see \citet{pesarin1992} and \citet{basso2009}]. A $p$-value for the
overall null hypothesis (that all individual null hypotheses are true)
is based either on the distribution of the resulting global statistic
$T$ or on a permutation approximation. If the unweighted sum combining
function is applied to the proposed~$T_i$, then
\begin{eqnarray*}
T&=&\int_{ [a, b  ]}\sum_{j=1}^k
n_j\bigl(\hat{\mu}_j(t)-\hat{\mu }(t)
\bigr)^2\,dt/(k-1)
\\
&=&\sum_{i=1}^m\int
_{ [a_i, b_i  ]}\sum_{j=1}^k
n_j\bigl(\hat{\mu }_j(t)-\hat{\mu}(t)
\bigr)^2\,dt/(k-1)\\
&=&\sum_{i=1}^mT_i.
\end{eqnarray*}

The closure procedure is then applied to perform the overall test based
on these combining functions as well as to adjust the individual
$p$-values for multiplicity. The closure method is based on testing all
nonempty intersections of the set of $m$ individual hypotheses, which
together form a closure set. The procedure rejects a given hypothesis
if all intersections of hypotheses that contain it as a component are
rejected. \citet{hochberg1987} show that the closure procedure controls
the family-wise error rate (FWER) at a strong level, meaning that the
type I error is controlled under any partial configuration of true and
false null hypotheses.

When the number of individual tests $m$ is relatively large, the use of
the closure method becomes computationally challenging. For example,
setting $m=15$ results in $2^{15}-1=32\mbox{,}767$ intersections of
hypotheses. \citet{hochberg1987} described a shortcut for the $T=\max\{
T_i\}$ combining function, where $T_i$ stands for the $i$th test
statistic for $i$ in the set of $H_i$ pertinent to a particular
intersection hypothesis. For this combining function they showed that
the significance for any given hypothesis in the closure set can be
determined using only $m$ individual tests. \citet{zaykin2002} described
a shortcut for the closure principle in the application of their
truncated $p$-value method (TPM) that uses an unweighted sum combining
function. In the next section we exploit the shortcut described by \citet
{zaykin2002} and show that for the $T=\sum T_i$ combining function the
required number of evaluations is $m(m+1)/2$.

%s2.3.1 #&#
\subsubsection{The shortcut version of the closure procedure}\label{sec2.3.1}

The shortcut version of the closure method for the unweighted sum
combining function should be implemented\vadjust{\goodbreak} as follows. First, order the
individual test statistics from minimum to maximum as $T_{(1)} \leq
T_{(2)} \leq\cdots\leq T_{(m)}$, where
%
%e4 #&#
\begin{equation}
\label{eq:Ti} T_i=\int_{ [a_i, b_i  ]}\sum
_{j=1}^k n_j\bigl(\hat{
\mu}_j(t)-\hat {\mu}(t)\bigr)^2\,dt/(k-1).
\end{equation}
Let $H_{(1)}, H_{(2)}, \ldots, H_{(m)}$ be the corresponding ordered
individual hypotheses of no difference among functional means on the
interval $ [a_{(i)}, b_{(i)}  ]$, $i=1, \ldots, m$. Now, among
intersection hypotheses of size two,
\begin{eqnarray*}
&T_{(1)}+T_{(2)}\leq T_{(1)}+T_{(3)}\leq
\cdots\leq T_{(1)}+T_{(m)},&
\\
&T_{(2)}+T_{(3)}\leq T_{(2)}+T_{(4)} \leq
\cdots\leq T_{(2)}+T_{(m)},&
\\
&\cdots&
\end{eqnarray*}
Here, the statistic $T_{(i)}+T_{(j)}$ corresponds to intersection
hypotheses $H_{(ij)}$ of no significant difference on both intervals
$ [a_{(i)}, b_{(i)} ] \cup [ a_{(j)}, b_{(j)} ]$.
Among intersections of size three,
\begin{eqnarray*}
&T_{(1)}+T_{(2)}+T_{(3)} \leq T_{(1)}+T_{(2)}+T_{(4)}
\leq\cdots\leq T_{(1)}+T_{(2)}+T_{(m)},&
\\
&T_{(2)}+T_{(3)}+T_{(4)} \leq T_{(2)}+T_{(3)}+T_{(5)}
\leq\cdots\leq T_{(2)}+T_{(3)}+T_{(m)},&
\\
&\cdots&
\end{eqnarray*}
Thus, significance for the hypothesis $H_{(m)}$ can be determined by
looking for the largest $p$-value among $m$ tests,
\[
T_{(m)}, T_{(m)}+T_{(1)}, \ldots, \sum
_{i=1}^mT_{(i)}.
\]
For the hypothesis $H_{(m-1)}$, the significance can be determined by
investigating the $p$-values corresponding to $(m-1)$ tests
\[
T_{(m-1)}, T_{(m-1)}+T_{(1)}, \ldots, \sum
_{i=1}^{m-1}T_{(i)},
\]
along with the $p$-value for the test $\sum_{i=1}^mT_{(i)}$ which is
already found. Finally, for the first ordered hypothesis $H_{(1)}$, the
significance can be determined by evaluating a single test $T_{(1)}$
and then looking for the largest $p$-value among it and the $p$-values
of the hypotheses $H_{(12)}$, $H_{(123)}, \ldots, H_{(12\cdots m)}$,
which are already evaluated. Thus, significance of any individual
hypothesis $H_{(i)}$ is determined using $m$ $p$-values, but the number
of unique evaluations to consider is $m+(m-1)+\cdots+1=m(m+1)/2$.

The described shortcut assumes that all distributions corresponding to
the test statistics are the same and the magnitude of the test
statistic has a monotonic relationship with its $p$-value. If the
$p$-values for the individual tests are determined from permutational
distributions (as in our situation), a bias will be introduced. The
bias is caused by a mismatch between the minimum value of the test
statistics and the maximum $p$-value. That is, the minimum statistic is
not guaranteed to correspond to the maximum $p$-value. The procedure
becomes liberal since the individual $p$-values are not always adjusted
adequately. To reduce and possibly eliminate the bias, we made the
following adjustment to the shortcut. First, we adjusted the individual
$p$-values according to the shortcut protocol described above and
obtained a set of adjusted individual $p$-values, $p_1,p_2, \ldots,
p_m$. Then, we ordered the individual test statistics based on the
ordering of the unadjusted individual $p$-values. That is, we order the
unadjusted $p$-values from maximum to minimum and get a corresponding
ordering of the test statistics\vspace*{1pt} $T_{(1)}^*,T_{(2)}^* ,\ldots, T_{(m)}^*$.
Now the inequality $T^*_{(1)} \leq T_{(2)}^* \leq\cdots\leq
T_{(m)}^*$ will not necessarily hold. We applied the shortcut based on
this new ordering and obtained another set of adjusted individual
$p$-values, $p_1^*, p_2^*, \ldots, p_m^*$. Finally, the adjusted
individual $p$-values were computed as $\max\{p_i, p_i^*\}, i=1, \ldots,
m$. This correction to the shortcut increases the number of the
required computations by a factor of two but still is of the order
$m^2$ instead of $2^m$.

A small simulation study was used to check whether this version of the
correction provides results comparable to adjustments generated by the
entire set of intersection hypotheses. For the four multiplicity
adjustment schemes: (i) correction based on the ordered test statistics
shortcut, (ii) correction based on the ordered unadjusted $p$-values
shortcut, (iii)\vadjust{\goodbreak} correction based on $\max\{p_i, p_i^*\}$ [combination of
both corrections (i) and (ii)], and (iv) the full closure method, we
obtained $p$-values under the global null based on 1000 permutations,
$m=5$, and conducted 1000 simulations, providing 5000 corrected
$p$-values. First, we were interested in how many times the $p$-values
adjusted by various shortcuts were ``underestimated'' (not corrected
enough) relative to the full closure method. The $p$-values adjusted by
a shortcut based on the ordered test statistics, $p_1, p_2, \ldots,
p_m$, were underestimated 554 out of 5000 times. The $p$-values
adjusted by a shortcut based on the ordered unadjusted $p$-values,
$p^*_1, p^*_2, \ldots, p^*_m$, were underestimated 60 out of 5000
times. The $p$-values adjusted using both corrections, $\max\{ p_i,p^*_i
\}$, $i=1,\ldots, m$, were underestimated 38 out of 5000 times. Second,
we compared type I error rates under the $\max\{ p_i,p^*_i \}$ shortcut
and the full closure method and found that they were \textit{exactly}
the same. The above results allowed us to conclude that the
multiplicity adjustment based on the $\max\{ p_i,p^*_i \}$ shortcut is adequate.

%f2 #&#
\begin{figure}

\includegraphics{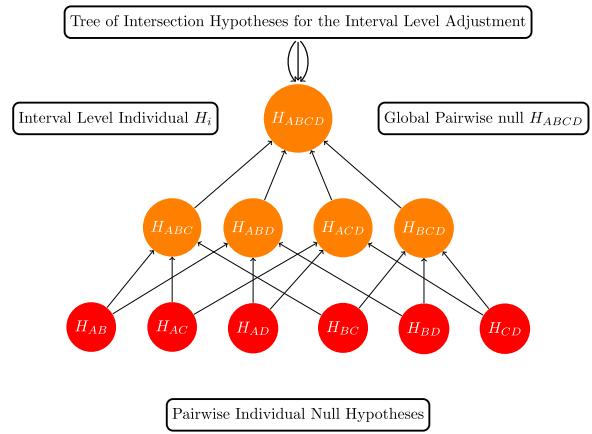}

\caption{Example of the closure set for the pairwise comparison of four groups.}
\label{fig:diag}
\end{figure}

%s2.3.2 #&#
\subsubsection{Proposed methodology for pairwise comparison of
functional means}

Above, we provided details on how to implement the proposed methodology
to isolate regions of the functional domain with statistically
significant differences and showed that with a computational shortcut
the closed testing scheme is computable even for a large number of
individual tests $m$. Now, we show how to further use the proposed
methodology to find pairs of functional means that are different within
the regions where statistical significance was identified. The
procedure is implemented as follows:
\begin{longlist}[(iii)]
\item[(i)] Within an interval $ [ a_i, b_i  ]$ with a
statistically significant difference among functional means, set the
$p$-value for the ``global'' null of no difference among functional
means to the adjusted individual $p$-value corresponding to that interval.
\item[(ii)] Compute the pairwise statistic as well as statistics for the
intersection hypotheses as in (\ref{eq:Ti}).
\item[(iii)] Find the $p$-values based on the permutation algorithm and
adjust them using the closure principle.
\end{longlist}
Figure~\ref{fig:diag} illustrates the closure set for pairwise
comparison of four populations. The $p$-value of the top node
hypothesis, $H_{ABCD}$, of no significant difference among the four
population means would be set equal to the adjusted $p$-value of the
interval level individual hypothesis of interest $H_i$, $i=1, \ldots,
m$. The bottom node individual hypotheses, $H_{AB}, \ldots, H_{CD}$,
correspond to no significant pairwise difference between groups $AB$,
$AC,\ldots,CD$ in this interval. Note that now the indexing of the
hypotheses corresponds to \textit{population means} instead of \textit
{intervals in the functional domain}.
The closure principle is used to adjust the individual $p$-values.

Certain issues may arise with a test of pairwise comparisons conducted
by global randomization. \citet{petrondas1983} noted that for the
overall equality hypothesis all permutations are assumed to be equally
probable, that is, the exchangeability among all treatment groups is
assumed. However, for the hypothesis of equality of a particular subset
of treatments, the global permutation distribution cannot be used
because differences in variability among the treatment groups can cause
bias in the statistical tests. The results of the simulation study,
presented in the next section, did not reveal any noticeable bias in
the permutation test. In the case of the pairwise comparison, our
method maintained good control of the type I error rate as well as had
enough power to correctly identify groups of unequal treatments. The
minimal bias observed might be due to a relatively small (three) number
of treatments that we chose to consider in our simulation study. \citet
{petrondas1983} and \citet{troendle2011} provide ways to perform
permutation tests correctly in the case of the pairwise comparison. We
leave implementation of these solutions for future research.

%s3 #&#
\section{Simulations}\label{s:sim}

Before proceeding to the description of our simulation study, we would
like to note that all functional data methods, including the one
proposed in this article, are affected by how well the estimated
functions approximate data. A failure to adequately approximate data
with smooth functions may result in a loss of statistical power. An
``adequate'' approximation is a subjective decision, however, below we
outline some choices that are intended to aid fitting particular data
at hand.

%s3.1 #&#
\subsection{Estimation of functional responses}\label{s:est}

Use of functional data methods requires a ``guess'' of a function, $\mu
(t)$, underlying each response. Since this function is generally
unknown, nonparametric methods are used to approximate it.
Nonparametric methods represent a function as a linear combination of
$K$ ``basis functions.'' A potential shortcoming of all testing
procedures based on nonparametric methods is ambiguity in the choice of
basis functions (e.g., splines, Fourier series, Legendre polynomials,
etc.) and the number of basis functions, $K$. An incautious choice
might lead to over- or under-fit and the resulting loss of statistical power.

The current consensus regarding the choice of basis functions,
supported, among others, by \citet{horvath2012}, \citet{storey2005}, \citet
{ramsay2005}, is that a good choice should mimic the general features
of the data. Specifically, the Fourier basis is recommended for
periodic, or nearly periodic, data and the $B$-spline basis for
nonperiodic locally smooth data. Since it is known that hemolytic
responses have a smooth ``S'' shape, the $B$-spline basis was a natural
choice in our application.

\citet{rice2001} and \citet{griswold2008} investigated the impact of the
number of basis functions, $K$, on the quality of fit to the data. More
specifically, \citet{rice2001} showed that the result of a functional
fit is rather insensitive to the specification of the number of basis
functions for the $B$-spline basis. \citet{griswold2008} provided general
recommendations for the number of basis functions with an arbitrary
basis. They showed that if the data result from (i) an erratically
changing stochastic process or (ii) a smoothly varying process with a
small measurement error, the recommended number of basis terms required
to fit the data is close to the number of observations per subject. We
chose the number of basis functions to be close to the number of
observations which coincides with the recommendations provided by \citet
{griswold2008}. During the course of the novocaine experiment the
percent of hemolysis was obtained by converting the spectrophotometric
readings. These readings---the measurements of the spectral
transmittance---were made with a spectrophotometer PE-5400 VI, which
has a low measurement error of $\pm0.5\%$ (details of the registration
certificate are at \url
{http://www.promecolab.ru/images/stories/Spektr/5400b-5400UF.pdf}).
Thus, we had an underlying process that is smooth with a small
measurement error so a higher number of basis functions was an
appropriate choice.

%s3.2 #&#
\subsection{Simulations setup}
\begin{figure}

\includegraphics{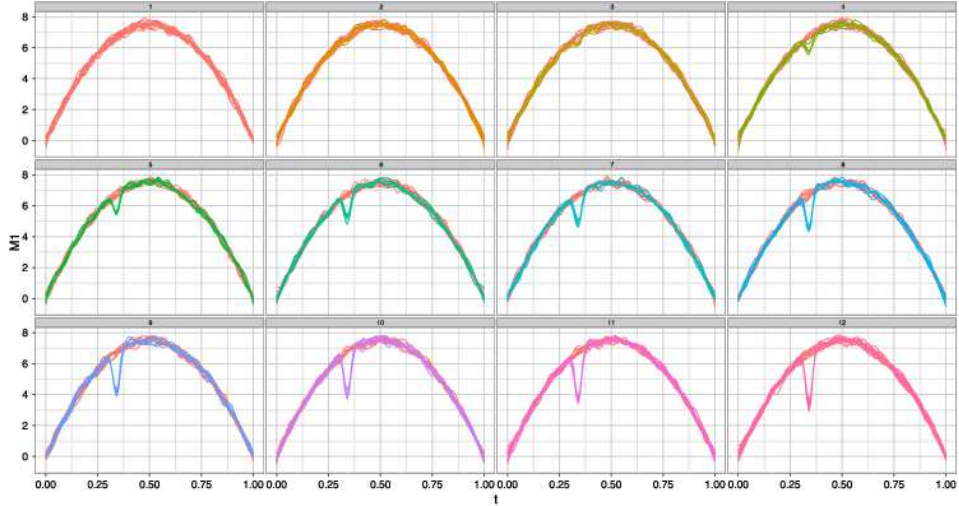}

\caption{Plot of the data for one simulation replicate under the M1
case and 12 different values of $\beta$.
The functions that have the mean value, $\mu_3(t)$, deviating from the
overall mean ($\mu_1(t)=\mu_2(t)\equiv\mu(t)$) are highlighted in a color.}
\label{fig:M1}\vspace*{-1pt}
\end{figure}

\begin{figure}[b]\vspace*{-1pt}

\includegraphics{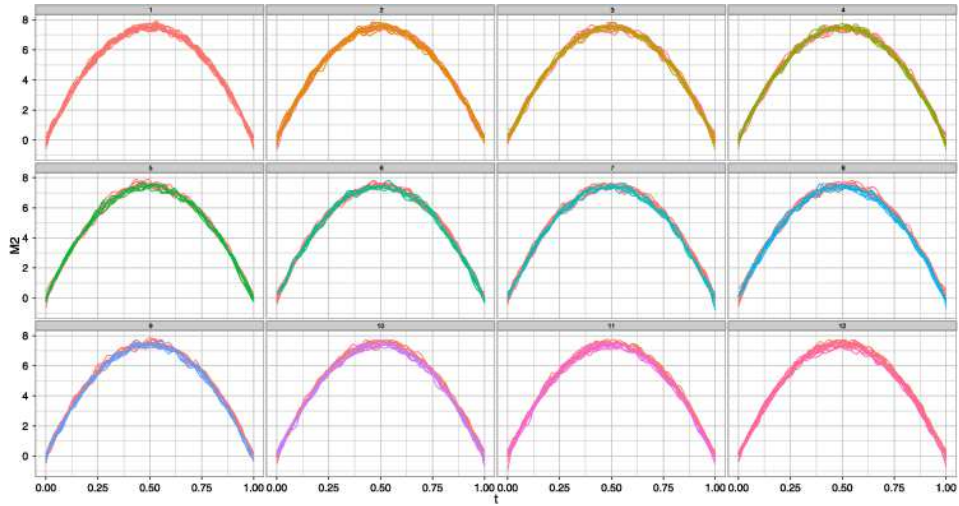}

\caption{Plot of the data for one simulation replicate under the M2
case and 12 different values of~$\beta$.
The functions that have the mean value deviating from the overall mean
are highlighted in a color.}
\label{fig:M2}
\end{figure}

Now, we describe a simulation study that we carried out in order to
evaluate the performance of our approach. A nonparametric fit to the
data was achieved by employing the $B$-spline basis functions with a
``knot'' at each observation over $t$. The number of basis functions,
$K$, is equal to the number of knots plus two. The simulations
scenarios were inspired by a Monte Carlo study in \citet{cuesta2010}. We
considered
\begin{longlist}[(M1)]
\item[(M1)] $f_i(t)=30(1-t)t-3\beta|\sin(16\pi t)|I_{\{0.325<t<0.3575 \}
}+\varepsilon_i(t)$,
\item[(M2)] $f_i(t)=30(1-t)t-\beta|\sin(\pi t/4)|+\varepsilon_i(t)$,
\end{longlist}
where $t_1=0, \ldots, t_{101}=1$, $\beta\in\{0.000, 0.045, 0.091,
0.136, 0.182, 0.227, 0.273,\break   0.318, 0.364, 0.409, 0.455, 0.500\}$,
and random errors $\varepsilon_i(t)$ are independently normally
distributed with mean zero and variance 0.3. Case M1 (illustrated in
Figure~\ref{fig:M1}) corresponds to a situation where a small set of
observations was generated under $H_A$ to create a spike. In M2
(illustrated in Figure~\ref{fig:M2}), a large number of observations
were generated under $H_A$ but the differences are less apparent [a
deviation along the entire range of $t$ that gradually increases from
$\min(t)$ to $\max(t)$]. The parameter $\beta$ controls the strength of
the deviation from the global null. The reason for considering these
two cases was to check the performance of our method for different
ranges of false null hypotheses.

%f3 #&#

%f4 #&#

In each case (M1 and M2), we generated three samples of functional data
with 5 observations from each group. The first two samples had the same
mean ($\beta=0$) and the third sample's mean was deviating ($\beta\neq
0$). Once the functional data were generated for different values of
$\beta\neq0$, we split the functional domain into different numbers
of equal-length intervals ($m=5$ and $m=10$) and evaluated the power of
rejecting the null hypotheses $H_0\dvtx \mu_1(t)=\mu_2(t)=\mu_3(t)$ at the
5\% level. We used 1000 simulations to obtain a set of power values for
each combination of $\beta$ and $m$ values. We used a permutation
test\vadjust{\goodbreak}
to obtain the $p$-values. This was achieved by randomly permuting the
original observations for each $t$ across groups 1000 times, and for
each new grouping, refitting the functional means and recalculating the
value of the test statistic. The $p$-value was found as the proportion
of 1000 recalculated test statistics greater than the observed statistic.

%s3.3 #&#
\subsection{Simulation results}
\begin{figure}

\includegraphics{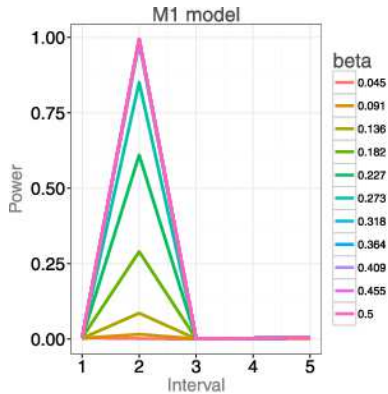}

\caption{The probability of rejecting the null hypothesis $H_0\dvtx \mu
_1(t)=\mu_2(t)=\mu_3(t)$ for $m=5$ intervals.}
\label{fig:5spike}
\end{figure}

Figure~\ref{fig:5spike} presents results of power evaluation for model
M1 and five intervals ($m=5$). Under this model, a set of observations
generated under $H_A$ fell into the second interval. That is, the
functional mean of the third sample had a spike deviation from the
functional mean of the first two samples over the second interval. The
magnitude of the spike increased monotonically as a function of $\beta
$. The plot shows that the proportion of rejections reveals a peak over
the region of the true deviation, while being conservative over the
locations with no deviations. Thus, we conclude that the proposed
methodology provides satisfactory power over the region with true
differences, while being conservative over the regions where the null
hypothesis is true.

%f5 #&#

\begin{figure}

\includegraphics{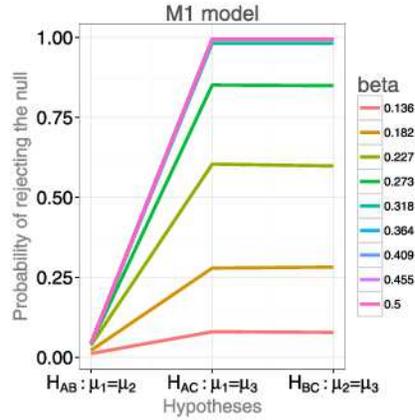}

\caption{The probability of rejecting individual pairwise hypotheses
$H_{AB}\dvtx \mu_1(t)=\mu_2(t)$, $H_{AC}\dvtx \mu_1(t)=\mu_3(t)$, and $H_{BC}\dvtx \mu
_2(t)=\mu_3(t)$.}
\label{fig:5spikepair}
\end{figure}

Once we identified the region of the functional domain with differences
in means (i.e., the second interval), we used the extension of the
proposed methodology to perform a pairwise comparison and determine
which populations are different. Figure~\ref{fig:5spikepair} provides
the results of power evaluation of the pairwise comparisons at the 5\%
significance level. In the case of $H_{AB}$ (where the null $\mu_1=\mu
_2$ is true), the simulation output tells us that the procedure is a
bit conservative, maintaining the type I error rate right below the 5\%
level for the higher values of $\beta$. In the case of $H_{AC}$ and
$H_{BC}$ (where the null is false), it can be seen that the power of
the pairwise comparison is satisfactory.

%f6 #&#

%t1 #&#
\begin{table}[b]
\caption{Power of the pairwise comparison assuming common means $\mu_1$
and $\mu_2$ over the 1st interval}
\label{t:5gradpair1}
\begin{tabular*}{\textwidth}{@{\extracolsep{\fill}}lccc@{}}
\hline
$\bolds{\beta}$ & $\bolds{H_{AB}\dvtx \mu_1=\mu_2}$ &
$\bolds{H_{AC}\dvtx \mu_1=\mu_3}$ & \multicolumn{1}{c@{}}{$\bolds{H_{BC}\dvtx \mu
_2=\mu_3}$} \\
\hline
0.318 & 0.027 & 0.021 & 0.026 \\
0.364 & 0.029 & 0.024 & 0.028 \\
0.409 & 0.031 & 0.034 & 0.038 \\
0.455 & 0.036 & 0.041 & 0.047 \\
0.500 & 0.036 & 0.049 & 0.054 \\
\hline
\end{tabular*}
\end{table}
%

%t2 #&#
\begin{table}[b]
\caption{Power of the pairwise comparison assuming common means $\mu_1$
and $\mu_2$ over the 2nd interval}
\label{t:5gradpair2}
\begin{tabular*}{\textwidth}{@{\extracolsep{\fill}}lccc@{}}
\hline
$\bolds{\beta}$ & $\bolds{H_{AB}\dvtx \mu_1=\mu_2}$ &
$\bolds{H_{AC}\dvtx \mu_1=\mu_3}$ & $\bolds{H_{BC}\dvtx \mu
_2=\mu_3}$ \\
\hline
0.273 & 0.018 & 0.049 & 0.057 \\
0.318 & 0.025 & 0.074 & 0.086 \\
0.364 & 0.031 & 0.104 & 0.116 \\
0.409 & 0.037 & 0.145 & 0.164 \\
0.455 & 0.041 & 0.214 & 0.224 \\
0.500 & 0.045 & 0.298 & 0.323 \\
\hline
\end{tabular*}
\end{table}

\begin{figure}

\includegraphics{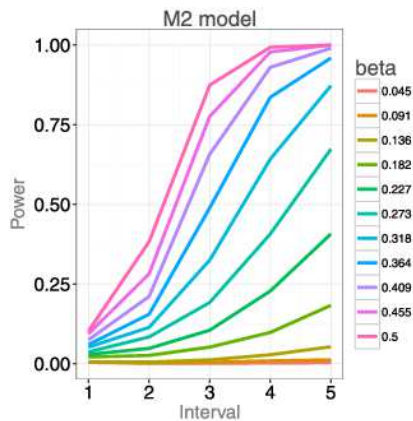}

\caption{The probability of rejecting the null hypothesis $H_0\dvtx \mu
_1(t)=\mu_2(t)=\mu_3(t)$ in the case of the M2 model and 5 intervals.}
\label{fig:5grad}\vspace*{-3pt}
\end{figure}

The results for the M2 case, where the number of true effects is large
and the magnitude of the effect gradually increases from $\min(t)$ to
$\max(t)$, are provided in Tables~\ref{t:5gradpair1}--\ref{t:5gradpair5}
and Figure~\ref{fig:5grad}. The plot shows that for a fixed value $\beta
$, the proportion of rejections of the hypothesis $H_0\dvtx \mu_1(t)=\mu
_2(t)=\mu_3(t)$ gradually increases with the magnitude of the effect.
Across different values of $\beta$, power values are also increasing,
attaining the value of 1 for the fifth interval and $\beta=0.5$. The
results of the pairwise comparisons are provided in
Tables~\ref{t:5gradpair1}--\ref{t:5gradpair5}. Power is the highest for the highest
value of $\beta$ (0.5), but overall the method does a good job of
picking out the differences between $\mu_1$ and $\mu_3$, and $\mu_2$
and $\mu_3$, while maintaining control of spurious rejections for $\mu
_1$ and $\mu_2$.

%t3 #&#
\begin{table}[t!]
\caption{Power of the pairwise comparison assuming common means $\mu_1$
and $\mu_2$ over the 3rd interval}
\label{t:5gradpair3}
\begin{tabular*}{\textwidth}{@{\extracolsep{\fill}}lccc@{}}
\hline
$\bolds{\beta}$ & $\bolds{H_{AB}\dvtx \mu_1=\mu_2}$ &
$\bolds{H_{AC}\dvtx \mu_1=\mu_3}$ & $\bolds{H_{BC}\dvtx \mu
_2=\mu_3}$ \\
\hline
0.182 & 0.015 & 0.038 & 0.040 \\
0.227 & 0.021 & 0.077 & 0.084 \\
0.273 & 0.027 & 0.160 & 0.155 \\
0.318 & 0.037 & 0.289 & 0.275 \\
0.364 & 0.041 & 0.437 & 0.434 \\
0.409 & 0.048 & 0.610 & 0.600 \\
0.455 & 0.048 & 0.731 & 0.735 \\
0.500 & 0.049 & 0.839 & 0.835 \\
\hline
\end{tabular*}
\end{table}
%

%t4 #&#
\begin{table}[t!]
\caption{Power of the pairwise comparison assuming common means $\mu_1$
and $\mu_2$ over the 4th interval}
\label{t:5gradpair4}
\begin{tabular*}{\textwidth}{@{\extracolsep{\fill}}lccc@{}}
\hline
$\bolds{\beta}$ & $\bolds{H_{AB}\dvtx \mu_1=\mu_2}$ &
$\bolds{H_{AC}\dvtx \mu_1=\mu_3}$ & \multicolumn{1}{c@{}}{$\bolds{H_{BC}\dvtx \mu
_2=\mu_3}$} \\
\hline
0.182 & 0.017 & 0.082 & 0.080 \\
0.227 & 0.023 & 0.207 & 0.196 \\
0.273& 0.030 & 0.375 & 0.365 \\
0.318 & 0.036 & 0.618 & 0.611 \\
0.364 & 0.039 & 0.817 & 0.807 \\
0.409 & 0.041 & 0.920 & 0.915 \\
0.455 & 0.041 & 0.971 & 0.971 \\
0.500 & 0.041 & 0.993 & 0.993 \\
\hline
\end{tabular*}
\end{table}

%t5 #&#
\begin{table}[t!]
\caption{Power of the pairwise comparison assuming common means $\mu_1$
and $\mu_2$ over the 5th interval}
\label{t:5gradpair5}
\begin{tabular*}{\textwidth}{@{\extracolsep{\fill}}lccc@{}}
\hline
$\bolds{\beta}$ & $\bolds{H_{AB}\dvtx \mu_1=\mu_2}$ &
$\bolds{H_{AC}\dvtx \mu_1=\mu_3}$ & $\bolds{H_{BC}\dvtx \mu
_2=\mu_3}$ \\
\hline
0.136 & 0.012 & 0.044 & 0.042 \\
0.182 & 0.020 & 0.164 & 0.160 \\
0.227 & 0.030 & 0.380 & 0.383 \\
0.273 & 0.038 & 0.640 & 0.645 \\
0.318 & 0.041 & 0.858 & 0.859 \\
0.364 & 0.042 & 0.955 & 0.957 \\
0.409 & 0.042 & 0.986 & 0.988 \\
0.455 & 0.042 & 0.997 & 1.000 \\
0.500 & 0.042 & 1.000 & 1.000 \\
\hline
\end{tabular*}  \vspace*{12pt}
\end{table}

Results based on $m=10$ intervals are similar to those based on $m=5$
intervals and can be found in the supplementary material
[Vsevolozhskaya, Greenwood and Holodov (\citeyear{vsevolozhskaya2014})]. A careful consideration of these results,
however, reveals that the procedure tends to lose power as the number
of intervals increases but gains power as the number of curves per
group increases.\looseness=-1

%f7 #&#

%s4 #&#
\section{Analysis of hemolysis curves}

In this section we illustrate the proposed methodology by applying it
to a study of the effect of novocaine conducted by \citet
{holodov2012}. The motivation behind the study was to investigate
pharmaceutical means of preventing the formation of stomach erosive and
ulcerative lesions caused by a long-term use of nonsteroidal
anti-inflammatory drugs (NSAIDs). Internal use of a novocaine solution
was proposed as a preventative treatment for NSAID-dependent complications.

During the course of the experiment, blood was drawn from male rats to
obtain an erythrocyte suspension. Then, four different treatments were
applied: control, low ($4.9\times10^{-6}$ mol/L), medium ($1.0 \times
10^{-5}$ mol/L), and high ($2.01\times10^{-5}$ mol/L) dosages of
procaine. After treatment application, the erythrocyte suspension was
incubated for 0, 15, 30, 60, 120 or 240 minutes. At the end of each
incubation period, hemolysis was initiated by adding 0.1 M of
hydrochloric acid to the erythrocyte suspension. The percent of
hemolysis or the percent of red blood cells that had broken down was
measured every 15 seconds for 12 minutes. The experiment was repeated 5
times for each dosage/incubation combination using different rats.
Therefore, the data set consists of 120 separate runs with 49
discretized observations per run and involves four experimental
conditions with six incubation times, replicated 5 times for each
treatment/incubation combination. For more details see
\citet{holodov2012}.\vadjust{\goodbreak}

%f8 #&#
\begin{figure}

\includegraphics{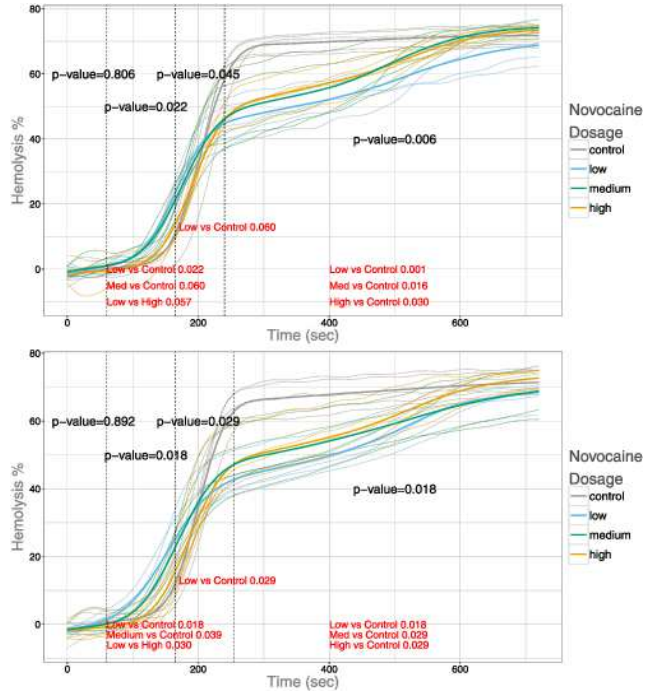}

\caption{Erythrogram means for the control group and the treatment
groups for 15 (top graph) and~30 (bottom graph) minute incubation times.}
\label{fig:means}
\end{figure}

We fit the data with smoothing cubic $B$-splines with 49 equally spaced
knots at times $t_1=0,\ldots,t_{49}=720$ seconds to generate the
functional data. The reasoning behind these choices is provided in
Section~\ref{s:est}. A smoothing parameter was selected by generalized
cross-validation (GCV) for each functional observation with an
increased penalty for each effective degree of freedom in the GCV, as
recommended in \citet{mgcv}. % Figure \ref{fig:means} shows the average
%curves for the control group and the treatment groups for the six
%incubation times.

To keep the analysis as simple as possible, each incubation data set
was analyzed for treatment effects separately. Our initial test was to
check for a significant difference in mean erythrograms (mean hemolysis
curves) anywhere in time among novocaine dosages. A Bonferroni
correction was applied to these initial $p$-values to adjust for
multiplicity at this level. The results indicated strong evidence of
differences for the 15 and 30 minute incubation times
($p$-value$_{\mathrm{Bonf}}=0.006$ and $p$-value$_{\mathrm{Bonf}}=0.018$, resp.).
Figure~\ref{fig:means} illustrates the results for these incubation
times. For the rest of the incubation times, we found no evidence
against the null hypothesis that the four erythrogram means coincided,
so no further analysis was conducted.

Next, we examined the 15 and 30 minute incubation results in more
detail to asses the nature of the differences. For both incubation
times, four time intervals of interest were prespecified: (i) the
latent period (0--60 sec), (ii)~hemolysis of the population of the least
stable red blood cells (61--165 sec), (iii) hemolysis of the general red
blood cell population (166--240 sec), and (iv) the plateau (over 240
sec). The latent period is associated with erythrocytes spherulation
and occurs between addition of the hemolytic agent and initiation of
hemolysis. The names of the next two periods are self-explanatory. The
plateau period is associated with deterioration of the population of
the most stable erythrocytes.

We applied our method to determine if statistical significance is
present in each of the four time intervals. In the application of our
method, we set the $p$-values for the global hypotheses $H_{1234}$ of
no significant difference on all four intervals to the Bonferroni
adjusted $p$-values obtained on the previous step. For the 15 minute
incubation time, no statistical significance was found during the
latent period ($p$-value${}=0.806$), and statistically significant results
were found during hemolysis of the least stable red blood cell
population ($p$-value${}=0.022$), general red blood cell population
(marginal significance with the $p$-value${}=0.060$) and plateau
($p$-value${}=0.006$). The same results were obtained from the 30 minute
incubation, that is, no statistical significance during the latent
period ($p$-value${}=0.892$) and statistical significance for the rest of
the time intervals with $p$-values of 0.018, 0.029 and 0.018 for the
periods of hemolysis of the least stable population, general population
and plateau, respectively.

Finally, we were interested in pairwise comparison of treatment levels
within the time intervals of statistical significance. Once again,
similar results were found for both incubation times, although the
$p$-values were often larger for the 15 minute incubation time. During
the hemolysis of the least stable red blood cell population, at least
some evidence was found of a difference between low dosage and control
($p$-value$_{15}=0.020$, $p$-value$_{30}=0.018$), medium dosage and
control ($p$-value$_{15}=0.060$, $p$-value$_{30}=0.039$), and low
dosage and high dosage ($p$-value$_{15}=0.057$,
$p$-value$_{30}=0.030$). During the hemolysis of the general
population, at least some evidence of a significant difference was
found between the low dose and control ($p$-value$_{15}=0.060$,
$p$-value$_{30}=0.029$). During the plateau interval, there was a
significant difference between low dose and control
($p$-value$_{15}=0.001$, $p$-value$_{30}=0.018$), medium dose and
control ($p$-value$_{15}=0.016$, $p$-value$_{30}=0.029$), and high dose
and control ($p$-value$_{15}=0.030$, $p$-value$_{30}=0.029$).

The results of the analysis can be summarized as follows. The rate of
hemolysis increases with the dosage of novocaine. That is, the
structural and functional modifications in the erythrocyte's membrane
induced by novocaine are dosage dependent. The results also indicate
the distribution of erythrocytes into subpopulations with low, medium
and high resistance to hemolysis. These populations modified by
novocaine react differently with the hemolytic agent. After 15 and 30
minutes of incubation, the ``old'' erythrocytes (least stable) modified
by low ($4.9\times10^{-6}$ mol/L) and medium ($1.0 \times10^{-5}$
mol/L) doses of procaine react faster to the hemolytic agent than those
under the control or the high ($2.01\times10^{-5}$ mol/L) dose.
However, reaction of the general and ``young'' (most stable)
erythrocyte population modified by the same (low and medium) dosages is
characterized by higher stability of the membrane and thus have higher
resistance to the hemolytic agent. Thus, novocaine in low and medium
doses has a protective effect on the general and ``young'' erythrocyte
populations. However, an increase in procaine dosage does not lead to
an increase of erythrocyte resistance to the hemolytic agent. The
effect of the high dose of novocaine ($2.01\times10^{-5}$ mol/L) does
not differ significantly from the control and thus is destructive
rather than protective.

Conclusions of our statistical analysis confirm certain findings
reported in a patent by \citet{holodov2012}. Specifically, our analysis
confirms that novocaine in low dosages tends to have a protective
effect. However, \citet{holodov2012} reported a significant difference
among erythrograms for all incubation times but zero minutes. This
inconsistency is due to a failure to properly adjust the number of
tests performed in the original analysis. The findings reported in the
current paper have a higher assurance that a replication experiment
will be able to detect the same differences reported here.

%s5 #&#
\section{Discussion}

We have suggested a procedure which allows researchers to find regions
of significant difference in the domain of functional responses as well
as to determine which populations are different over these regions. To
the best of our knowledge, there are no existing competing procedures
to the proposed methodology. Thus, our numerical results reported in
Section~\ref{s:sim} do not include a comparison of the proposed method
to other alternatives. Nevertheless, the simulations revealed that our
procedure has satisfactory power and does a good job of picking out the
differences between population means. Also, in our simulation study, a
relatively small number of regions ($m=5$ and $m=10$) were considered.
A higher number of individual tests (intervals) can be easily
implemented with the described shortcut to the closure principle.

The relative efficiency of all nonparametric functional approaches
depends on the ``adequate'' representation of data by smooth functions.
In Section~\ref{s:est} we provided some general recommendations that
should help a reader to choose an effective basis and a number of basis
functions for a particular application. A valid point raised by one of
the reviewers was that if power of any function-valued statistical
procedure depends on the accuracy of the estimates of individual
curves, they, in turn, might depend on the number of observed time
points per subject. \citet{griswold2008} studied this issue and showed
that as the number of measurements per subject increased from 10 to 20,
the power of a functional approach remained relatively constant or
improved.\vadjust{\goodbreak} \citet{berk2011} used as little as 10 observations per subject
to estimate the functional responses. Thus, we expect statistical power
to be rather insensitive to the number of time points at hand as long
as researchers have at least 10 observations and are producing
reasonable functional estimates.

Another important issue is that the nonparametric approaches based on
the $B$-spline basis might suffer from a phenomenon termed ``edge
effect''---a bias in the estimation at the endpoints. Thus, power of
the procedure to detect differences among functional responses might be
affected at the intervals near the edges if the estimated smooth
functions have boundary artifacts (e.g., unexpected behavior). This was
not the case in our simulation study nor in our application. If a
researcher encounters functional boundary artifacts while fitting
particular data of interest, s/he might consider correcting for this
effect [e.g., see \citet{masri2004}].

We also note that for the procedure presented in this article, the
regions of interest in the functional domain should be prespecified
prior to the analysis. However, in our experience researchers have
never had a problem with {a priori} region identification. From
previous research, expected results as well as specific regions of
interest are typically known. We also mention that in the application
of our method the intervals should be mutually exclusive and
exhaustive. If researchers are interested in a test over overlapping
intervals, the solution is to split the functional domain into smaller
mutually exclusive intervals for individual tests (terminal nodes of
the hypotheses tree). The decision for the overlapping region would be
provided by a test of an intersection hypothesis (``higher'' node in
the hypotheses tree). We also expect the intervals to be exhaustive
since it would be unexpected for researchers to collect data over time
periods that they have no interest in. Finally, if for some reason
distinct regions cannot be prespecified, a large number of equal sized
intervals can easily be employed, however, this might result in loss of power.

The present work has two open issues that suggest a direction for
future research. First, the method is conservative and so a more
powerful approach may be possible. Second, the permutation strategy for
the pairwise comparison test may lead to biased inference. Solutions to
the latter problem were suggested both by \citet{petrondas1983} and \citet
{troendle2011}. We leave implementation of these solutions for future
research, as this seems to be a minor issue with a small number of
treatment groups as are most often encountered in FANOVA applications.

\begin{supplement}[id=suppA]
\stitle{Additional simulation results}
\slink[doi]{10.1214/14-AOAS723SUPP} %[doi,text={...}] - jei reikia suskaldyti doi
\sdatatype{.pdf}
\sfilename{aoas723\_supp.pdf}
\sdescription{Additional simulation results for the two models (M1 or
M2), two different number of intervals ($m=5$ or $m=10$), and either 5
or 20 subjects per group are summarized in the tables below. Overall,
these results indicate that the procedure tends to lose power as the
number of intervals increases but gains power as the number of subjects
per group increases.}
\end{supplement}

% imsref loaded by akundreckaite, 2014-03-10 09:26:42
% imsref loaded by akundreckaite, 2014-03-10 09:59:58

\printaddresses

\end{document}